\documentclass[sigconf]{acmart}

\usepackage{algorithm}
\usepackage{algorithmicx}
\usepackage{algpseudocode}
\usepackage{CJK}
\usepackage{graphicx}
\usepackage{subfigure} 
\usepackage{array}
\usepackage{amsmath}
\usepackage{amsmath,bm}
\usepackage{soul}
\usepackage{multirow}
\AtBeginDocument{%
  \providecommand\BibTeX{{%
    \normalfont B\kern-0.5em{\scshape i\kern-0.25em b}\kern-0.8em\TeX}}}

\acmConference[SenSys '21]{SenSys '21: ACM Conference on Embedded Networked Sensor Systems}

\begin{document}

\title{Detecting and Correcting IMU Movements During Joint Angle Estimation}


\author{Chunzhi Yi}

\affiliation{%
  \institution{Harbin Institute of Technology}
  \streetaddress{92 West Dazhi Str.}
  \city{Harbin}
  \state{Heilongjiang}
  \country{China}
  \postcode{150001}
}
\email{chunzhiyi@hit.edu.cn}

\author{Feng Jiang}
\authornotemark[1]
\affiliation{%
  \institution{Harbin Institute of Technology}
  \streetaddress{92 West Dazhi Str.}
  \city{Harbin}
  \state{Heilongjiang}
  \country{China}
  \postcode{150001}}
\email{fjiang@hit.edu.cn}

\author{Baichun Wei}
\affiliation{%
  \institution{Harbin Institute of Technology}
  \streetaddress{92 West Dazhi Str.}
  \city{Harbin}
  \state{Heilongjiang}
  \country{China}
  \postcode{150001}}

\author{Chifu Yang}
\affiliation{%
  \institution{Harbin Institute of Technology}
  \streetaddress{92 West Dazhi Str.}
  \city{Harbin}
  \state{Heilongjiang}
  \country{China}
  \postcode{150001}}

\author{Zhen Ding}
\affiliation{%
  \institution{Harbin Institute of Technology}
  \streetaddress{92 West Dazhi Str.}
  \city{Harbin}
  \state{Heilongjiang}
  \country{China}
  \postcode{150001}}

\author{Jubo Jin }
\authornotemark[1]
\affiliation{%
  \institution{Heilongjiang University of Finance and Economic}
  \streetaddress{Xueyuan Ld.}
  \city{Harbin}
  \state{Heilongjiang}
  \country{China}
  \postcode{150001}}
\email{jinjubo98@163.com}

\author{Jie Liu}
\affiliation{%
  \institution{Harbin Institute of Technology}
  \streetaddress{Taoyuan Str.}
  \city{Shenzhen}
  \state{Guangdong}
  \country{China}
  \postcode{518055}}
\email{jieliu@hit.edu.cn}

\renewcommand{\shortauthors}{C.Yi, et al.}

\begin{abstract}
 Inertial measurement units (IMUs) increasingly function as a basic component of wearable sensor network (WSN)systems.  IMU-based joint angle estimation (JAE) is a relatively typical usage of IMUs, with extensive applications. However, the issue that IMUs move with respect to their original placement during JAE is still a research gap, and limits the robustness of deploying the technique in real-world application scenarios. In this study, we propose to detect and correct the IMU movement online in a relatively computationally lightweight manner. Particularly, we first experimentally investigate the influence of IMU movements. Second, we design the metrics for detecting IMU movements by mathematically formulating how the IMU movement affects the IMU measurements. Third, we determine the optimal thresholds of metrics by synthetic IMU data from a significantly amended simulation model. Finally, a correction method is proposed to correct the effects of IMU movements. We demonstrate our method on both synthetic data and real-user data. The results demonstrate our method is a promising solution to detecting and correcting IMU movements during JAE.
\end{abstract}



\maketitle

\section{Introduction}
Wearable sensor networks (WSN), as a key component of the human-centered Internet-of-Things (IoT), have been widely used for health monitoring, risk and ability assessing and home automation \cite{8060403, yang2006body}. Among the various wearable sensors contained in the WSN, the inertial sensors, i.e.  the inertial measurement units (IMUs), provide the information of body motions with relatively robust performance and low cost, and increasingly serve as a basic configuration of many WSN applications \cite{9122466,lara2012survey}. One example of the IMU-based body motion tracking is to estimate joint angles. IMU-based joint angle estimation (JAE) is a relatively mature technique with extensive applications, such as take-home rehabilitation \cite{ibrahim2020inertial}, robotics \cite{9075258} and sports risk assessment \cite{fasel2017validation}. Comparing with other joint angle estimation techniques like the optical motion capture system or the mechanical motion capture system, the IMU-based technique benefits from its wearable and small-volume characteristics, thus can enable an easy-to-integrate and low-cost usage \cite{chen2016toward}.

An IMU includes a three-axis accelerometer, a three-axis gyroscope and optionally a three-axis magnetometer, providing linear accelerations, angular rates and local magnetic fields in the form of three-dimensional vectors. Using IMU measurements to estimate joint angles generally relates to the following three steps. First, the so-called “absolute orientations” or “attitudes” of IMUs can be estimated to track the relative orientational relationship between the IMU coordinate frames and the Earth coordinate frame during body motions \cite{yi2018estimating, ligorio2015novel}. In practice, the attitudes of IMUs cannot be used directly to estimate joint angles, because the IMU coordinate frames are not aligned to the coordinate frames of body segments or joints. This misalignment would cause a large estimating error. This is the reason of performing the calibration procedures, which are to align the IMU coordinate frames with the body coordinate frames.  The calibration procedures can be performed via specific body postures \cite{roetenberg2009xsens}, customized calibration devices \cite{hagemeister2005reproducible,adamowicz2019validation} or some kinematic constraints \cite{seel2014imu,yi2019sensor, 9116999}. Specific body postures or customized calibration devices should be performed before joint angle estimation. Developing the alignment using kinematic constraint is to construct a cost function by assuming the existence of some fixed joint axes or joint centres. It can be performed during JAE  using the IMU measurements in a buffer. Some of the kinematic constraint-related methods can even avoid the estimation of IMUs’ attitudes \cite{seel2014imu,9116999}. Finally, the rotational relationship between the body frames of two adjacent body segments can be obtained by multiplying the rotation matrices estimated by the first two steps. Then, joint angles can be estimated by decomposing the rotational relationship.

An implicit assumption of the currently used methods is that the IMUs will not move with respect to their mounted body segments. And the calibration procedures shall only be performed once. However, if the IMUs moved during JAE due to some occasional collisions, large inertia or loose attachment, a large estimating error would be induced. The solution is to perform the calibration procedures again and to re-establish the IMU-to-body alignment. For the specific body postures or the calibration devices, the online joint angle estimation should be terminated in order to perform the calibration again. For the kinematic constraint-related methods, we proposed to maintain a running buffer to update the estimates of joint axes constantly, thus to update the IMU-to-body alignment \cite{9116999}. As presented in our previous study \cite{9116999}, the joint angles can be estimated constantly using the output of the last running buffer. Due to the gradient-based optimization, the approach would use relatively high computational resource thus might impede the real-time and online detection of IMU movements. We aim to minimize the IMU movements’ effects on the online JAE and avoid to interrupt the online estimation progress as far as possible. Thus, we focus on detecting the timing of the IMU movement and then correcting the consequent error by re-establishing the IMU-to-body alignment in a new buffer. And we will compare it with our previously used method in this paper. 

We aim to develop an online detection and correction algorithm for IMU movements, which works in an easy-to-use manner without consuming too much computational resource. The design principle of the algorithm is 1) to measure the IMU movement  \textit{sensitively}, and 2) not to cause too many misdetections, i.e.  \textit{robustness}. To this end, we analyze what metrics can be used to detect the movements, how sensitive the metrics can be, and how the experimental paradigm can be designed to evaluate the effectiveness of the detection algorithm. Based on the analysis, we propose to calculate the metrics in sliding windows and detect the IMU movement if the metrics exceed a threshold. Then, the errors caused by the IMU movement can be corrected simply by restarting a buffer to estimate the joint axes. The \textbf{key challenges} can be summarized as follows.

\begin{itemize}
\item The \textit{metrics} of IMU movements: One of the most intuitive metrics is the difference of the estimated joint axes between sliding windows. The joint axis is the coordinate axis of the body coordinate frame described in the IMU coordinate frame. If IMU moves, the changes of the estimated joint axis’ coordinates will be the direct consequence of the IMU movement. However, it might not be the optimal metric.  Keeping estimating joint axes in sliding windows has to perform gradient-based optimizations in each sliding window. The consequent computational time might impede the real-time detection of IMU movements. An alternative is to directly use the measurements of IMUs, i.e. the linear accelerations and angular rates, as metrics. Other than the IMU movement-caused variation, the difference of the measurements between sliding windows also contains the variations caused by the body segment’s motions and measuring noise. This might make the IMU movement-caused variation overwhelmed by the variations caused by the factors except for the IMU movement. Thus, the metrics that use IMU measurements should be designed carefully. The relationship between the IMU measurements and the IMU movements’ effects should be studied, aiming to instruct the design of appropriate metrics.
\item The \textit{threshold} of detecting IMU movements: To make the detection algorithm both sensitive and robust is to balance the tradeoff between the possibility of misdetections and the minimum magnitude of the detected IMU movement. That is, other than designing appropriate metrics, optimal thresholds of the employed metrics should also be determined to detect the minimum IMU movement and to accommodate the metrics’ variations caused by other factors like body motions and measuring noise. The data that covers various conditions under IMU movements should be leveraged to determine the optimal threshold. And multiple factors should be considered. For example, if we use the estimated joint axes’ coordinates as the metric, determining the threshold should consider the estimation error and the IMU movement-caused variations of the coordinates. If the IMU measurements are used as metrics, the variations caused by body motions and measuring noise should also be considered. 
\item The \textit{simulation model} for IMU movements: As stated above, a relatively large amount of data under various conditions of IMU movements, body motions and IMU-to-body attachments should be collected to determine the threshold and to evaluate the efficiency of the detection algorithm. It would be labor expensive to collect the data in real-user experiments to include various magnitudes and directions of IMU movements. Moreover, due to the irregular shape of body segments and the deformable soft tissues of human, the magnitudes and directions of IMU movements would be hard to control and measure. To solve this issue, a reasonable and efficient simulation model should be proposed to synthesize the data under IMU movements during locomotion. 
\end{itemize}

Aiming to solve the above mentioned challenges, we take an initial trial to study some basic metrics of IMU movements and to develop a detection and correction method. This proof-of-concept study focuses on the lower-limb angle estimation, due to its relatively extensive researches and wide applications. We leave the extension to upper-limb joints in our future work. Our \textbf{key contributions} are as follows.
\begin{itemize}
\item To the best of our knowledge, this is the first study that investigates the IMU movement issue of joint angle estimation. 
\item We mathematically formulate the relationship between IMU measurements and the effects of IMU movements, and design the metrics of IMU movements accordingly.
\item We propose a reasonable simulation model to synthesize data of IMU movements and use it to determine the optimal threshold for the detection algorithm.
\item Based on the designed metrics and optimal thresholds, we demonstrate a convenient-to-use and computationally lightweight method to detect IMU movements and correct the consequent errors.
\end{itemize}

\section{Related Works}
Before diving into our method, we would like to indicate some experience that we learn from the literature and leverage in our study.

Studies on \textbf{IMU movement} mainly focus on human activity recognition (HAR). K. Kunze et al. \cite{6926690} categorized  the possible IMU movements that might occur in HAR. According their category scheme, considerable studies focused on the "displacement within a body part", which aimed to find an IMU placement and orientation-independent feature set or classifier \cite{banos2014dealing,alinia2015impact}. In this way, the algorithm can perform HAR without limiting the specific placements of an IMU with respect to its mounted body segment. Some other studies focused on the "on-body placement", which proposed to identify the body location the IMUs placed at \cite{alinia2015impact}. In so doing, body-location-specific or body-location-free algorithms can be performed to let HAR accommodate more locations \cite{alinia2015impact, rokni2018autonomous}. Due to the difference tasks, we focus on different aspects. If speaking with their language, our scope is more like the "displacement within a body part". We study IMU movements during JAE within a body segment. Compared with HAR, there is no "on-body placement" issue in JAE, since IMUs must be placed on the body segments beside a joint of interest. And we do not focus on displacement variations like putting IMUs in a pocket. Because JAE is with more specific and professional usages than HAR, it would be hard to find a scenario to place IMUs in a pocket and to calculate joint angles.

To the best of our knowledge, the study that relates to our scope the most is \cite{brennan2011assessment} that studied the anatomical frame variation effect on JAE. Compared with it, our study focuses on how sensor movement affects the measurements of sensors rather than the results of estimating joint angles. On investigating IMU movements, our study can be seen as the preliminary of \cite{brennan2011assessment}. Moreover, our study enables the real-time detection of IMU movements.

\textbf{Metrics} function as a key in detecting IMU movements. If considering the issue in a anomaly detection perspective, we aim to find a way to detect the “collective anomalies” \cite{erhan2020smart} and to accommodate measuring noise and unideal conditions such that misdetections can be suppressed. In this initial study, we would like to investigate the metrics, i.e. what values obtained from the measurements are most suitable for detecting IMU movements. With the studied metrics, various anomaly detection algorithms could be easily integrated to achieve a better performance in the future study.

As stated before, we leverage a \textbf{simulation model} to determine the optimal thresholds. We use virtual simulation models rather than mechanical gimbals. Mechanical gimbals benefit from its flat surface, known geometric parameters and easy-to-control motions, were extensively used in JAE-related studies \cite{brennan2011quantification,yi2021reference}. However, the optimal thresholds of metrics would be significantly influenced by the characteristics of human motions. It would be tricky for gimbals to mimic human motions well. Virtual simulation models could generate IMU movements from scratch \cite{seel2012joint} or from the data obtained by optical motion capture system \cite{young2011imusim} or even from video \cite{kwon2020imutube,rey2019let}. Considering our emphasis on human motions, we choose to synthesize IMU measurements from the data obtained by optical motion capture system using the model presented in \cite{young2011imusim} and propose to amend it for synthesizing data under IMU movements.

\section{Preliminaries}

In this section, we show how the IMU movements affect the accuracy of JAE, and introduce some basic background information of JAE.

\subsection{The Effect of IMU Movements}
\textbf{Set-up:} One subject (male, 23 years old, 178cm, 65kg) was asked to walk on a treadmill with his self-selected speed. As shown in Fig., two IMUs (Trigno Wireless system; DELSYS, Boston, MA, USA, 148.148Hz) were put on the subject’s thigh and shank. We used the optical motion capture system as the reference of joint angles. The system details are presented in section 5. During walking, the IMU mounted on the shank was manually rotated, i.e. IMU movement was induced. 
\begin{figure}[h]
  \centering
  \includegraphics[width=\linewidth]{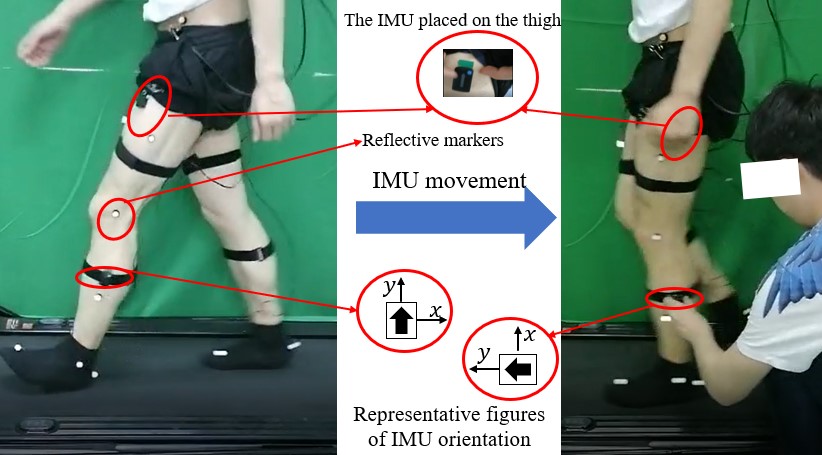}
  \caption{The schematic diagram of inducing IMU movements.}
\end{figure}

\textbf{Performance analysis:} We estimate the knee angles in the sagittal plane using the measurements of the two IMUs. The employed algorithm is the work presented in \cite{seel2014imu}. The initial 2000 sample points during walking are used to estimate the joint axes. We denote the estimated knee angles before and after moving the IMU on the shank by \textit{normal} and \textit{moved}. 

\begin{figure}[h]
  \centering
  \includegraphics[width=3cm]{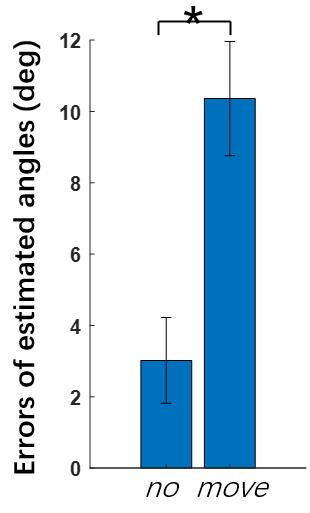}
  \caption{The errors of estimated knee angles without IMU movement ($no$) and after an IMU movement ($move$).$*$ denotes the significant difference of the metric compared with other metrics (ANOVA, p < 0.050).}
  \label{move}
\end{figure}

We present the estimated knee angles' errors of normal and moved. The one-way ANOVA is used to analyze the statistical significance.  It can be seen in Fig. \ref{move} that the knee angle estimation error under the condition normal is similar to the error reported in \cite{seel2014imu}.It can be seen in Fig. \ref{move} that the errors caused by the IMU movement are significantly large, which cause an obvious deviation of estimates. 

Based on the performance, we could conclude that the IMU movement may induce a dramatic error increase. This might further affect the consequent usages of JAE, since estimated joint angles might contribute to controlling robots or helping to make some clinical decisions. Thus, the detection of IMU movements shall be timely. Furthermore, considering JAE is often performed in edge nodes, the detection algorithm should be computationally lightweight.

\subsection{Terminology Definition}
\begin{figure}[h]
  \centering
  \includegraphics[width=6cm]{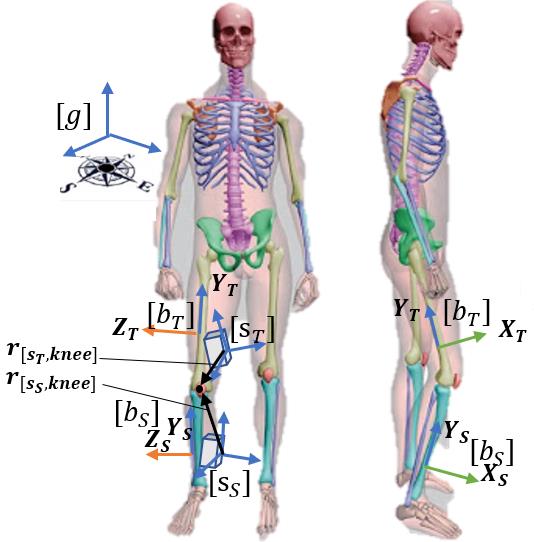}
  \caption{The schematic diagram of coordinate systems and vectors.}
  \label{coordinate}
\end{figure}

In the following, we describe the coordinate systems and the terminologies we use in our study. As shown in Tab. \ref{tab:term} , we define the following terminologies and signs to denote the terms we use. As shown in Fig. \ref{coordinate}, we use the following coordinate systems.
\begin{itemize}
\item The global coordinate frame $[g]$: The internal frame whose axes correspond to the gravity and the magnetic field.  
\item The body coordinate frame $[b_i]$: The coordinate frame fixed on body segment. The axes and the rotations between adjacent frames are determined by anatomy \cite{wu2002isb}.
\item The IMU coordinate frame $[s_i]$: The coordinate frame fixed on an IMU.The IMU measurements are described in the coordinate frame.
\end{itemize}

 \begin{table*}
  \caption{The Terminologies}
  \label{tab:term}
  \begin{tabular}{cc}
    \toprule
    Terminologies & Meanings\\
    \midrule
    $\bm{x}$ & A vector\\
	$\bm{\hat{x}}$ &  A vector after an IMU movement\\
	$\bm{\Delta x}$ & $\bm{\hat{x}}$  - $\bm{x}$ \\
	$\bm{x^{[f]}}$ & The vector $\bm{x}$ described in the coordinate frame $[f]$ \\
	$(\bm{x})_{window}$ & The vector $\bm{x}$ in a sliding window \\
	$\textit{R}$ & A rotation matrix\\
	$\textit{R}^{[f1]}_{[f2]}$ & The rotation matrix denoting the rotation from the coordinate frame $[f1]$ to the coordinate frame $[f2]$\\
	$\textit{R}_{J_k}(t)$ & The rotation matrix denoting the rotation of the joint $J_k$ at time t\\
	$\bm{a_i}(t)$ & The measured acceleration of the $i$th body segment at time  t, $i = T,S, denoting thigh and shank$  \\
	$\bm{\omega_i}(t)$ &  The measured angular rate of the $i$th body segment at time  t \\
	$\bm{g}(t)$ & The measured gravity acceleration at  time t\\
	$\bm{j_{J_k}}$ & The real value of the joint axis $j$ of  the $k$th joint $J_k$ \\
	$\bm{\tilde{j}_{J_k}}$ & The estimate of the joint axis $j$ of the $k$th joint $J_k$  \\
	$\theta_{J_k}(t)$ & The joint angle of the $k$th joint $J_k$ at the time instance \\
	$[s_i]$ & The IMU coordinate frame mounted on the $i$th body segment\\
	$[g]$ & The global coordinate frame\\
	$[b_i]$ & The body coordinate frame of the the $i$th body segment\\ 
	$\bm{r_{[s_i],J_k}}$ & The vector from the origin of the IMU coordinate system to the rotation center of joint $J_k$ \\
  \bottomrule
\end{tabular}
\end{table*}

\section{The 3-DoF Simulation Model For IMU Movements}

In this section, we describe the simulation model that synthesizes IMU measurements of two adjacent body segments linked by a 3-DoF joint. Also, the IMU movement model is included. As stated above, the simulation model leverages the biomechanical principles of joints and the kinematics measured by optical motion capture system, so as to well incorporate the characteristics of human body motions. In this section, without the loss of generality, we assume 1) two IMUs are placed on the thigh and the shank, respectively, in order to estimate knee angles; 2) the IMU on the thigh is moved.

\subsection{Data Preparation}

As stated above, in order to incorporate the characteristics of body motions, we choose to synthesize IMU measurements on the basis of real-user data. We use the optical motion capture system (MoCap) to collect the real-user data, i.e. the three-dimensional angles, orientations and positions of body segments and joints with respect to the global coordinate frame. The details of the system and the experiments are stated in section 5.

\subsection{The Simulation Model}
As stated in section 2, we use the data from optical motion capture system to synthesize the  IMU measurements mainly based on the IMUSim \cite{young2011imusim}. In our study, we set hip as the base of forward kinematics and use IMUSim to derive the orientations and positions of each limb and joint. We just focus on the measurements of accelerometer and gyroscope. Moreover, we amend some details to the algorithm of IMUSim in order to adapt it to our paradigm. 
\begin{itemize}
\item We synthesize arbitrary orientations between body coordinate frames and the global coordinate frame in each simulation run. As shown in line 2 of Algorithm \ref{IMUsim}, IMUSim leverages the orientation between the body coordinate frame on the hip and the global coordinate frame of the MoCap system.  This might limit variety of the synthesized data. 
\item We synthesize arbitrary orientations between body-fixed coordinate frames and IMU coordinate frames in each simulation run. As shown in line 6 of the algorithm, IMUSim leverages the orientation built by the single standing calibration procedure, which may just provide one kind of IMU-to-body alignment. This might also limit variety of the generated data.
\item We synthesize the data under IMU movements.
\end{itemize} 

\begin{algorithm}
\caption{The Algorithm of IMUSim \cite{young2011imusim}}
\label{IMUsim}
\begin{algorithmic}[1] 
\Require Input: data from MoCap , StartTime, EndTime
\State Smoothing and interpolating the data from MoCap
\State  $\bm{a_{i}^{[g]}}(t), \bm{\omega_i^{[g]}}(t), \bm{r_{[s_i],J_k}^{[g]}}, \textit{R}_{J_k}(t)$ $\gets$ Forward kinematics using the data from MoCap from StartTime to EndTime 
\State Adding effects of mechanical and electrical components of sensors by subsystem modeling
\State Adding effects of wireless communication
\end{algorithmic}
\end{algorithm}

\begin{algorithm}
\caption{Our Amended Algorithm For Simulation}
\label{amended}
\begin{algorithmic}[1] 
\Require data from MoCap , window, interval, StartTime, EndTime
\For {$\Phi_{mag} = 0 \to \pi$}
\For {$SimNum = 0 \to 1000$}
\State $\textit{R}_{[g]}^{[bT]}$, $\textit{R}_{[b_i]}^{[s_i]}$ $\gets$ Generate RanRot
\State $\bm{r_{[sT],knee}^{[g]}}$, $\bm{r_{[sS],knee}^{[g]}}$ $\gets$ A unit vector $*$ length $\in$ $(0,0.3)$
\For {$t = StartTime \to EndTime$}
\State Smoothing and interpolating the data from MoCap
\State  Forward kinematics using the data from MoCap and $\bm{r_{[sT],knee}^{[g]}}$, $\bm{r_{[sS],knee}^{[g]}}$, $\textit{R}_{[g]}^{[bT]}$
\State Adding effects of mechanical and electrical components of sensors by subsystem modeling
\State Adding effects of wireless communication
\State Transforming signals into IMU coordinate frames using $R_{[b_i]}^{[s_i]}$
\EndFor
\State $T_{initial}$ $\in$ [0, EndTime-window-interval]
\State $\textit{R}_{[sT]}^{[sT']}$ , $\bm{r_{move}}$ $\gets$ Generate Movements
\State $(\bm{\omega_i^{[s_i]}})_{window1}$ = $\bm{\omega_i^{[s_i]}}(:, T_{initial}\to T_{initial}+window)$, $(\bm{a_i}^{[s_i]})_{window1}$ = $\bm{a_i^{[s_i]}}(:, T_{initial}\to T_{initial}+window)$
\State $(\bm{\tilde{j}_{knee}^{[s_T]}})_{window1}$,$(\bm{\tilde{j}_{knee}^{[s_S]}})_{window1}$ $\gets$ Estimate JointAxis
\State $(\bm{\hat{\omega}_i^{[s_i]}})_{window2}$, $(\bm{\hat{a}_i^{[s_i]}})_{window2}$ $\gets$ Generate MovementEffect
\State $(\bm{\tilde{j}_{knee}^{[sT]}})_{window2}$,$(\bm{\tilde{j}_{knee}^{[sS]}})_{window2}$ $\gets$ Estimate JointAxis
\State Calculating metrics
\EndFor
\EndFor
\end{algorithmic}
\end{algorithm}

As shown in Algorithm \ref{amended}, we develop a simulation model for  3-DoF lower-limb joints based on IMUSim \cite{young2011imusim}. The IMUs are assumed to be rigidly attached to the segments. The simulation model can be developed in the following steps. 

1. For each simulation run (i.e. each $SimNum$), as shown in line 2 and 3, we randomly synthesize the placements of IMUs.  Particularly, we randomly synthesize a rotation matrix $\textit{R}_{[g]}^{[bT]}$ to denote an arbitrary pose of the base of the forward kinematics, i.e. the hip.  Since the orientation between the body coordinate frame on the hip and the global coordinate frame of the MoCap system is assumed to hold in IMUSim. We synthesize various attitudes of  body coordinate frames by changing the orientation of the global frame (i.e. the matrix  $\textit{R}_{[g]}^{[bT]}$). The rotation matrix $\textit{R}_{[b_i]}^{[s_i]}$ is also randomly synthesized to denote arbitrary orientations between a body coordinate frame and the IMU coordinate frame placed on the body segment. As shown in Algorithm \ref{ranrot}, three unit vectors that are perpendicular to each other are randomly synthesized to construct the rotational matrix, $R_{[f1]}^{[f2]}$. The three vectors are to denote the three coordinate axes of $[f1]$ coordinate frame described in $[f2]$ coordinate frame \cite{diebel2006representing}. 

\begin{algorithm}
\caption{Generate RanRot}
\label{ranrot}
\begin{algorithmic}[1] 
\State $\bm{x}$,$\bm{y}$ $\gets$ random unit vectors
\State $\bm{z}$ = $\bm{x}\times \bm{y}$
\State $\bm{y}$ = $\bm{z}\times \bm{x}$
\State $R_{[f1]}^{[f2]}$ = [$\bm{x}$,$\bm{y}$,$\bm{z}$]
\end{algorithmic}
\end{algorithm}

Then, we randomly synthesize vectors $\bm{r_{[s_T],J_k}^{[g]}}$ and $\bm{r_{[s_S],J_k}^{[g]}}$ to denote the vectors from the origin of IMU coordinate frames to the origin of body coordinate frames. These vectors are to represent the translational placements of IMUs. 

 As shown in line 7 and 11 of Algorithm \ref{amended}, our synthesized rotations are applied on the forward kinematics and synthesized IMU measurements. As shown in Eq. (\ref{acc_syn}) and (\ref{omega+syn}), we raise thigh and hip as the example, and the accelerations and angular rates are obtained by multiplying the rotation matrices with the synthesized vectors.

\begin{equation}
\label{acc_syn}
  \bm{a_T^{[s_T]}} = \textit{R}_{[b_T]}^{[s_T]} \cdot \textit{R}_{[g]}^{[bT]} \cdot (\bm{a_{hip}^{[g]}} + \textit{R}_{hip} \cdot \bm{a_{T,motion}^{[g]}}) + \bm{\delta_a} 
\end{equation}
\begin{equation}
\nonumber \bm{a_{T,motion}^{[g]}} = \bm{\dot \omega_T^{[g]}} \times \bm{r_{[s_T],hip}^{[g]}}  + (\bm{r_{[s_T],hip}^{[g]}}  \cdot  \bm{\omega_T^{[g]}}) \cdot \bm{r_{[s_T],hip}^{[g]}}
\end{equation}

\begin{equation}
\label{omega+syn}
  \bm{\omega_T^{[s_T]}} = \textit{R}_{[b_T]}^{[s_T]} \cdot \textit{R}_{[g]}^{[bT]} \cdot \bm{\omega_T^{[g]}} + \bm{\delta_{\omega}}
\end{equation}

2. We synthesize the IMU movements’ effects on IMU measurements. We consider two consecutive windows. The first window contains the IMU measurements before the IMU movement, while the second one contains the IMU measurements after the IMU movement. In each simulation run, we first randomly generate an initial time instant $T_{initial}$ denoting the start time of the first window. As shown in lines 15 of Algorithm \ref{amended}, we then use the IMU measurements in the first window to estimate the joint axes $(\bm{\tilde{j}_{knee}^{[sT]}})_{window1}$ and $(\bm{\tilde{j}_{knee}^{[sS]}})_{window1}$. For the second window, IMU movements are synthesized and denoted by the rotation $\textit{R}_{[sT]}^{[sT']}$ and a translation $\bm{r_{move}}$. As shown in Algorithm \ref{movement}, the movement-caused rotation is denoted by three Euler angles, i.e. the three elements of $\bm{\Phi}$. The movement-caused translation is randomly synthesized to denote the translation of the origin of the IMU coordinate frame, which will induce the change of the body motion-caused linear accelerations. The detailed caculation is presented in Algorithm \ref{move_effect}.

\begin{algorithm}
\caption{Generate Movements}
\label{movement}
\begin{algorithmic}[1] 
\Require $\Phi_{mag}$
\State $\bm{\Phi}$ $\gets$ $\Phi_{mag}$ $*$ a random unit three-element vector
\State $\textit{R}_{[sT]}^{[sT']}$ $\gets$ generate a XYZ rotation according to $\bm{\Phi}$
\State $\bm{r_{move}}$  $\gets$  a random unit three-element vector $*$ $\alpha$, $\alpha \in (0,0.15)$
\end{algorithmic}
\end{algorithm}

\begin{algorithm}
\caption{Generate MovementEffect}
\label{move_effect}
\begin{algorithmic}[1] 
\Require $\textit{R}_{[s_T]}^{[s_T']}$, $\bm{r_{move}}$, $\bm{\omega_T^{[s_T}}$, $\bm{a_T^{[s_T]}}$, window, interval, $T_{initial}$ 
\For {$t = (T_{initial}+interval) \to (T_{initial}+interval+window)$}

\State $(\bm{\hat{\omega}_T^{[s_T']}})_{window2}(:,t) = \textit{R}_{[s_T]}^{[s_T']} \cdot (\bm{\omega_T^{[s_T]}})_{window2}(:,t)$ 
\State  $(\bm{\hat{\omega}_S^{[s_S]}})_{window2}(:,t) = (\bm{\omega_S^{[s_S]}})_{window2}(:,t)$
\State Calculating $\bm{\hat{a}_T^{[s_T']}}$ by multiplying $\textit{R}_{[s_T]}^{[s_T']}$ and adding $\bm{r_{move}}$ with $\bm{r_{[s_T],hip}^{[g]}}$ in Eq. \ref{acc_syn}
\State  $(\bm{\hat{a}_S^{[s_S]}})_{window2}(:,t) = (\bm{a_S^{[s_S]}})_{window2}(:,t)$
\EndFor
\end{algorithmic}
\end{algorithm}

As shown in line 2 and 4 of Algorithm \ref{move_effect}, the IMU movement-caused translation induces a change of the body-motion-caused acceleration. The IMU movement-caused rotation induces the measurements of both angular rates and linear accelerations. Finally, we use the synthesized measurements $(\bm{\hat{\omega}_T^{[s_T']}})_{window2}$,$(\bm{\hat{a}_S^{[s_S']}})_{window2}$ to estimate the joint axes $(\bm{\tilde{j}_{knee}^{[s_T]}})_{window2}$,$(\bm{\tilde{j}_{knee}^{[s_S]}})_{window2}$, using the method presented in \cite{yi2019sensor}.

3. We calculate and store the metrics in each simulation run. The metrics are calculated by the equations in section 5.2. Moreover, we also calculate the change of the estimated coordinates of joint axes ( $\bm{\Delta \tilde{j}_{knee}^{[s_i]}}$) and the error of estimating the joint axes ($\bm{error \tilde{j}_{knee}^{[s_i]}}$). $\bm{error \tilde{j}_{knee}^{[s_i]}}$ is given by:
\begin{equation}
\label{error}
\bm{error \tilde{j}_{knee}^{[s_i]}} = \bm{j_{knee}^{[s_i]}} - \bm{\tilde{j}_{knee}^{[s_i]}}
\end{equation}

\section{The Mathematical Formulation and The Metric Design}

In this section, we formulate the effects of IMU movements, i.e. how IMU movements affect IMU measurements. With this mathematical formulation, we design the metrics that can be used to detect IMU movements and determine the optimal threshold for the metrics through a greedy algorithm. 

\subsection{Mathematical Formulation }
In order to instruct the design of the metrics and the detection algorithm, we mathematically formulate the IMU movements’ effects on the IMU measurement. Essentially, the IMU movement relates to the change of the IMU-to-body alignment. That is, the IMU coordinate frame rotates with respect to the body coordinate frame. The coordinates of the joint axis described in the IMU coordinate frame change. We can simply claim that coordinates’ change of the joint axis represent the effects of the IMU movement. Thus, we formulate the changes of  IMU measurements and the coordinates of the joint axis if an IMU moves.

For lower-limb joints, there is always a main axis around which the joint rotates the most time during gait cycles. This rationalizes the assumption that lower-limb joints can be approximated as hinge joints \cite{yi2019sensor,seel2014imu,9116999}. Following this commonly adopted assumption, Eq. (\ref{function1}) holds for the angular rates and the joint axis.

\begin{equation}
\label{function1}
F1 = 1/2 \cdot \sum_t (\| \bm{\omega_T^{[s_T]}}(t) \times \bm{\tilde{j}_{knee}^{[s_T]}}\|_2^2 - \| \bm{\omega_S^{[s_S]}}(t) \times \bm{\tilde{j}_{knee}^{[s_S]}}\|_2^2) = 0 
\end{equation}

If the IMU mounted on the thigh is moved, both the measurements of the IMU and the joint axis’ coordinates described in the IMU coordinate frame will change. That is, the coordiantes of  $ \bm{\tilde{j}_{knee}^{[s_T]}}$ and $\bm{\omega_T^{[s_T]}}$ will change. If Eq. (\ref{function1})  holds, the consequent change of $F1$, denoted by $\Delta F1_{\bm{j^{[s_T]}}}$, will be counteracted by $F1$’s change caused by the change of $ \bm{\omega_T^{[s_T]}}$ ($\Delta F1_{\bm{\omega_T}}$). That is,
\begin{equation}
\label{delta_F1}
\Delta F1_{\bm{j^{[s_T]}}} + \Delta F1_{\bm{\omega_T}} = 0
\end{equation}

Following the Taylor series expansion, $\Delta F1_{\bm{\omega_T}}$ and $\Delta F1_{\bm{j^{[s_T]}}}$ can be approximately expressed as:

\begin{math}
\begin{aligned}
\Delta F1_{\bm{\omega_T}} &\approx     \sum_t \frac{\partial F1}{\partial \bm{\omega_T^{[s_T]}}(t) } \cdot \bm{ \Delta \omega_T^{[s_T]}}(t) \\
&= -\sum_t (\bm{\omega_T^{[s_T]}}(t))^T \cdot sin(\angle_{\omega,j}) \cdot \bm{ \Delta \omega_T^{[s_T]}}(t)
\end{aligned}
\end{math}

\begin{math}
\begin{aligned}
\Delta F1_{\bm{j^{[s_T]}}} &\approx     \sum_t \frac{\partial F1}{\partial\bm{\tilde{j}_{knee}^{[s_T]}}} \cdot \bm{ \Delta \tilde{j}_{knee}^{[s_T]}} \\
&= \sum_t \| \bm{\omega_T^{[s_T]}}(t)\|_2^2 \cdot sin(\angle_{\omega,j}) \cdot (\bm{\tilde{j}_{knee}^{[s_T]}})^T \cdot \bm{ \Delta \tilde{j}_{knee}^{[s_T]}} 
\end{aligned}
\end{math}

where $\angle_{\omega,j} = <\bm{\omega_T^{[s_T]}}(t),\bm{\tilde{j}_{knee}^{[s_T]}}>$, denoting  the angle between $\bm{\omega_T^{[s_T]}}(t)$ and $\bm{\tilde{j}_{knee}^{[s_T]}}>$, $\bm{ \Delta \omega_T^{[s_T]}}(t)$ and $\bm{ \Delta \tilde{j}_{knee}^{[s_T]}}$ denote the differences of angular rates and joint axis’ coordinates caused by the IMU movement, respectively. Substituting  $\Delta F1_{\bm{\omega_T}}$ and $\Delta F1_{\bm{j^{[s_T]}}}$ in to Eq.(\ref{delta_F1}), we can obtain
\begin{align}
\label{tuidao1}
\nonumber \sum_t (\bm{\omega_T^{[s_T]}}(t))^T \cdot &\bm{ \Delta \omega_T^{[s_T]}}(t) 
\\
&- \sum_t \|\bm{\omega_T^{[s_T]}}(t) \|_2^2 \cdot (\bm{\tilde{j}_{knee}^{[s_T]}})^T \cdot  \bm{ \Delta \tilde{j}_{knee}^{[s_T]}} = 0
\end{align}

We imagine the condition when the moved IMU moves back to its original orientation under the same body motion. The changes of the angular rates and the coordinates of the joint axis, denoted by  $\bm{ \Delta \hat{\omega}_T^{[s_T]} }(t)$ and $\bm{ \Delta \tilde{j}_{knee}^{[s_T']}}$, can be expressed as
\begin{equation}
\label{converse1}
\nonumber \bm{ \Delta \hat{\omega}_T^{[s_T]} }(t) = \bm{\omega_T^{[s_T]}}(t) - \bm{\hat{\omega}_T^{[s_T]}}(t) = -\bm{ \Delta \omega_T^{[s_T]} }(t)  
\end{equation}

\begin{equation}
\label{converse2}
\bm{ \Delta \tilde{j}_{knee}^{[s_T']}} = \bm{\tilde{j}_{knee}^{[s_T]}} - \bm{\tilde{j}_{knee}^{[s_T']}} = - \bm{ \Delta \tilde{j}_{knee}^{[s_T]}}
\end{equation}

Then, we go through again and do the same thing for $\bm{ \Delta \hat{\omega}_T^{[s_T]} }(t)$ and $\bm{ \Delta \tilde{j}_{knee}^{[s_T']}}$. We can obtain
\begin{align}
\label{tuidao2}
\nonumber  -\sum_t (\bm{\hat{\omega}_T^{[s_T]}}(t))^T \cdot  &\bm{ \Delta \hat{\omega}_T^{[s_T]} }(t) 
\\
&+  \sum_t  \| \bm{\hat{\omega}_T^{[s_T]}}(t) \|_2^2 \cdot (\bm{\tilde{j}_{knee}^{[s_T]}})^T  \cdot  \bm{ \Delta \tilde{j}_{knee}^{[s_T']}} = 0  
\end{align}

Because $\bm{ \Delta \omega_T^{[s_T]} }(t)$ and $\bm{ \Delta \hat{\omega}_T^{[s_T]} }(t)$ have the same magnitude, we can get the following equation by summing Eq. (\ref{tuidao1}) and Eq.(\ref{tuidao2}).
\begin{equation}
\label{results} 
-\sum_t \|\bm{ \Delta \omega_T^{[s_T]} }(t) \|_2^2 + \sum_t \|\bm{ \omega_T^{[s_T]} }(t) \|_2^2 \cdot \|\bm{\Delta \tilde{j}_{knee}^{[s_T]}} \|_2^2= 0       
\end{equation}

Thus, the magnitude of the IMU movement-caused change of joint axis’ coordinates, i.e. the magnitude of $\bm{\Delta \tilde{j}_{knee}^{[s_T]}}$ , can be calculated by 
\begin{equation}
\label{results2} 
 \|\bm{\Delta \tilde{j}_{knee}^{[s_T]}} \|_2^2= \frac{\sum_t \|\bm{ \Delta \omega_T^{[s_T]} }(t) \|_2^2 }{\sum_t \|\bm{ \omega_T^{[s_T]} }(t) \|_2^2}      
\end{equation}

Eq.(\ref{results2}) suggests that the magnitude of IMU movement-caused changes relates to the changes of angular rates. That is, under the assumption we adopt, we obtain the magnitude of IMU movements’ effects analytically. 

We consider the linear accelerations. Under the hinge-joint assumption, we can get the following equation \cite{9116999}.
\begin{equation}
\label{function2}
F2 = ((\bm{a_T^{[s_T]}})^T \cdot  \bm{  \tilde{j}_{knee}^{[s_T]}})^2 - ((\bm{a_T^{[s_S]}})^T \cdot  \bm{  \tilde{j}_{knee}^{[s_S]}})^2 = 0
\end{equation}

Similarly, if the IMU mounted on the thigh is moved, the joint axis-caused change of $F2$ ( $\Delta F2_{\bm{j^{[s_T]}}}$) will be counteracted by the acceleration-caused change of $F2$ ($\Delta F1_{\bm{a_T}}$). And Eq. (\ref{delta_F2}) holds.
\begin{equation}
\label{delta_F2}
\Delta F2_{\bm{j^{[s_T]}}} + \Delta F2_{\bm{a_T}} = 0
\end{equation}

Then, the Taylor series expansion can be used to approximate $\Delta F2_{\bm{j^{[s_T]}}}$ and $\Delta F1_{\bm{a_T}}$.

\begin{math}
\label{tuidao_a1}
\Delta F1_{\bm{a_T}} \approx 2 \cdot \sum_t (\bm{a_T^{[s_T]}}(t))^T \cdot \bm{\tilde{j}_{knee}^{[s_T]}}) \cdot (\bm{\tilde{j}_{knee}^{[s_T]}})^T \cdot \bm{ \Delta a_T^{[s_T]} }(t)
\end{math}

\begin{math}
\label{tuidao_a2}
\Delta F2_{\bm{j^{[s_T]}}} \approx   2 \cdot \sum_t (\bm{a_T^{[s_T]}}(t))^T \cdot \bm{\tilde{j}_{knee}^{[s_T]}}) \cdot (\bm{\tilde{j}_{knee}^{[s_T]}})^T \cdot \bm{ \Delta \tilde{j}_{knee}^{[s_T]}}
\end{math}

Then, we get 
\begin{align}
\label{houxu}
\nonumber & \sum_t (\bm{a_T^{[s_T]}}(t))^T \cdot \bm{\tilde{j}_{knee}^{[s_T]}}) \cdot (\bm{\tilde{j}_{knee}^{[s_T]}})^T \cdot \bm{ \Delta a_T^{[s_T]} }(t)  
\\ 
&+  \sum_t (\bm{a_T^{[s_T]}}(t))^T \cdot \bm{\tilde{j}_{knee}^{[s_T]}}) \cdot (\bm{a_T^{[s_T]}}(t))^T \cdot \bm{ \Delta \tilde{j}_{knee}^{[s_S]}} = 0    
\end{align}                
                   
With the same operation, the relationship between  $\bm{ \Delta a_T^{[s_T]} }(t)$ and $\bm{ \Delta \tilde{j}_{knee}^{[s_T']}}$ can be obtained and summed with Eq. (\ref{houxu}), given by 
\begin{equation}
\label{retuls_a}
\sum_t (\bm{ \Delta \tilde{j}_{knee}^{[s_T]}})^T \cdot \bm{ \Delta a_T^{[s_T]} }(t) = 0  
\end{equation}

Eq. (\ref{retuls_a}) suggests that the direction of the IMU movement-caused changes relates to the changes of linear accelerations. That is, under the assumption we adopt, we obtain the direction of IMU movements’ effects analytically.

\subsection{Metric Design}
In the following, we consider how to design the metrics using the mathematical formulation. First, there are some limitations of the mathematical formulation that might influence the detection performance.
\begin{itemize}
\item \textit{Sliding window}: All the differences, i.e. $\bm{ \Delta \tilde{j}_{knee}^{[s_T]}}, \bm{ \Delta \omega_T^{[s_T]} }, \bm{ \Delta a_T^{[s_T]} }$, are calculated in the same sliding window. In the detection algorithm, we shall calculate such differences between difference sliding windows.

\item \textit{1-DoF assumption}: In the mathematical formulation, we take the hinge-joint assumption. For real lower-limb joints, there are rotations around other axes. 
\end{itemize}
Considering the limitations, we transform Eq.(\ref{results2}) into the following metrics to accommodate the condition of different sliding windows.

Metric1: 
\begin{align}
\nonumber mean(&(\frac{\sum_t \|\bm{ \Delta \omega_T^{[s_T]} }(t) \|_2^2 }{\sum_t \|\bm{ \omega_T^{[s_T]} }(t) \|_2^2})_{window1}, 
&(\frac{\sum_t \|\bm{ \Delta \omega_T^{[s_T]} }(t) \|_2^2 }{\sum_t \|\bm{ \omega_T^{[s_T]} }(t) \|_2^2})_{window2})
\end{align}

Metric2: 
\begin{math}
\frac{\sum_t \|\bm{ \Delta \omega_T^{[s_T]} }(t) \|_2^2 }{(\sum_t \|\bm{ \omega_T^{[s_T]} }(t) \|_2)_{window1} \cdot (\sum_t \|\bm{ \omega_T^{[s_T]} }(t) \|_2)_{window2}}
\end{math}

Similarly, we transform Eq.(\ref{retuls_a})into the following metrics.

Metric3:
\begin{math}
\| mean(\sum_t \bm{ \Delta a_T^{[s_T]} }(t))   \|
\end{math}

Metric4:
\begin{math}
\| mean(\sum_t \frac{ \bm{ \Delta a_T^{[s_T]} }(t)}{\| \bm{ \Delta a_T^{[s_T]} }(t) \|})   \|
\end{math}

Metric5:
\begin{align}
\nonumber \| mean(\sum_t (\frac{(\bm{a_T^{[s_T]}}(t))_{window1}}{\|(\bm{a_T^{[s_T]}}(t))_{window1}\|}
- \frac{(\bm{a_T^{[s_T]}}(t))_{window2}}{\|(\bm{a_T^{[s_T]}}(t))_{window2}\|})) \|
\end{align}

The magnitudes of $\bm{ \Delta a_T^{[s_T]} }(t)$ or the accelerations in different windows are divided, since the difference of accelerations only relates to the direction of IMU movements.

Moreover, the difference of the estimates of the joint axes is calculated as another metric, for a comparison purpose. The difference of the estimated joint axes, $\bm{ \Delta \tilde{j}_{knee}^{[s_T]}}$, is calculated as Metric6:
\begin{math}
\| (\bm{\tilde{j}_{knee}^{[s_T]}})_{window2}-(\bm{\tilde{j}_{knee}^{[s_T]}})_{window1} \|
\end{math}

\subsection{The Optimal Threshold For Each Metric}
In the following, we determine the thresholds for the metrics we design above. An ideal threshold is expected to 1) exclude the variations of normal measurements such that misdetection can be avoided, 2) include the abnormal measurements caused by the IMU movement such that a minimum IMU movement can be detected. With this in mind, we propose to use a greedy algorithm to search the optimal threshold for every metric we design and aim to check to what extent each metric can fulfil the expectations of an ideal threshold. 

\begin{algorithm}
\caption{A Greedy Method For finding the optimal threshold}
\label{threshold}
\begin{algorithmic}[1] 
\Require Metric$_i$ with $\Phi_{mag} = 0 \to \pi$
\State Initialize $\alpha = \max(Metric_i(\Phi_{mag}=0))$, $threshold_{normal} = 0.9, threshold_{moving} = 0.95, k = 0$
\State Calculating $r_{normal} = Number(Metric_i(\Phi_{mag}=0) <= \alpha )/ Number(Metric_i(\Phi_{mag}=0))$
\If  $r_{normal} \le threshold_{normal}$
\State Break the loop, go to line 7
\EndIf
\For $ \Phi \in$  $\{$ $\Phi$ | $\Phi_{mag}$ $\neq$ 0$\}$
\State k ++
\State $r_{moving}(k) = Number((Metric_i(\Phi)\geq \alpha )/ Number(metrics(\Phi))$
\EndFor
\State $\alpha = \alpha - 5e-3$, go to line 2
\State Output $\min_k(r_{moving}(k) \geq threshold_{moving}), \alpha+5e-3$
\end{algorithmic}
\end{algorithm}

As shown in Algorithm \ref{threshold}, we calculate the optimal threshold of each metric by detecting IMU movements with a minimum possibility of $0.95$ and maintaining a maximum misdetection possibility of $0.9$. We set $threshold_{normal}$ as the maximum misdetection possibility and $threshold_{moving}$ as the minimum detection possibility. $threshold_{normal}$ is set to be smaller than $threshold_{moving}$, because the cost of misdetection is just to restart a new buffer and to estimate the joint axes, which is smaller than the cost of not detecting an IMU movement. 

\section{ Experiments and Results}

The experiments are designed with the following aims: 1) demonstrating the simulation model of IMU movements, 2) evaluating the mathematical formulation presented in section 5.1, and 3) evaluating the designed metrics,4) evaluating the correction method for IMU movements. Since the magnitudes and orientations of IMU movements are hard to control in real-user experiments, we combine the simulation and the real-user experiments for the evaluations. For the hyper-parameters of our algorithm, we set $window$ as 2000s following the experience of our previous study, and set the lengths of the $interval$ between windows as 3000s and 5000s for a comparison purpose. The software we use to process data is MATLAB B 2015. 

\subsection{Data Collection}

\begin{figure}[h]
  \centering
  \includegraphics[width=\linewidth]{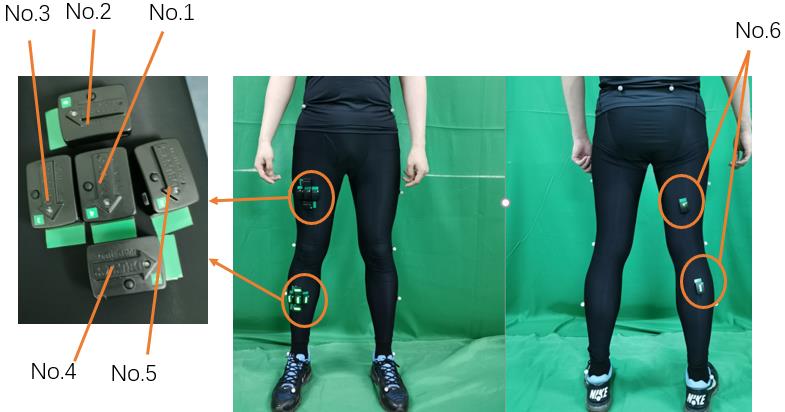}
  \caption{The schematic diagram of the experimental setup.}
  \label{experiment}
\end{figure}

Ten healthy subjects (7 males and 3 females, age range: 20-30 years, height range: 155-184cm, weight range: 50kg-90kg) were asked to walk with self-selected speeds to walk on the level ground. The experiment on each locomotion mode was repeated 3 times, each lasting about 2-3 min. The order of the task modes was randomly assigned. Rest periods were allowed between trials to avoid fatigue. In this experiment, we focused on the IMU movements on thigh and shank. As shown in Fig. \ref{experiment}, two sets of IMUs (Trigno Wireless system; DELSYS, Boston, MA, USA, 148.148Hz) were attached to subjects’ thigh and shank, each set consisting of six IMUs. As shown in Fig., No.1 - No.5 IMUs were attached closely, we used them to study the minimum movement we can induce in real-user experiments. No.6 was attached to the composite side of No.1 IMU with random orientations, in order to study the effect of a movement with a relatively large magnitude. Sixteen retro-reflective markers were attached to subjects’ pelvis and lower limbs following the principles of \cite{van2007color,leardini2011multi}. Both No.1 and No.6 IMUs were attached three additional markers, respectively, in order to obtain the positions and orientations of both IMUs \cite{young2010distributed}. The 3-D locations of the markers were recorded (100 Hz) using an 8-camera video system (Vicon, Oxford, UK). The joint angles were calculated by the pose estimation and inverse kinematics model embedded in the software Visual 3D. The signals from nine-axis IMUs and the video system were synchronized by the trigger and time stamps.

\subsection{Evaluation}

1. \textbf{Demonstrating the simulation model:} Following the method in \cite{young2011imusim}, we use the positions and orientations of No.1 and No.6 IMUs to calibrate the simulation model, respectively. In so doing, we can get simulated signals for both IMUs. We treat No.1 IMU as the original IMU and treat No.6 IMU as the consequence of IMU movement. That is, we assume No. 1 IMU is moved to the position and orientation of No. 6 IMU at a random instant. Specifically, for each subject, we first randomly generate a time instant $\tilde{t}$ as the timing of an IMU movement. We use the measurements of No1 IMU before $\tilde{t}$ and use the measurements of No.6 IMU after $\tilde{t}$. Then, we replace the relative position and orientation of both IMUs obtained by the MoCap system into our simulation model. We can simulate the IMU movement from No. 1 IMU to No.6 IMU. Finally, we evaluate our simulation model by comparing the simulated signals and the measurements of No.6 IMU.

\begin{figure}[ht]
\centering
\subfigure[Comparison of measured and simulated data without IMU movement.]{
\includegraphics[width = 0.4\textwidth]{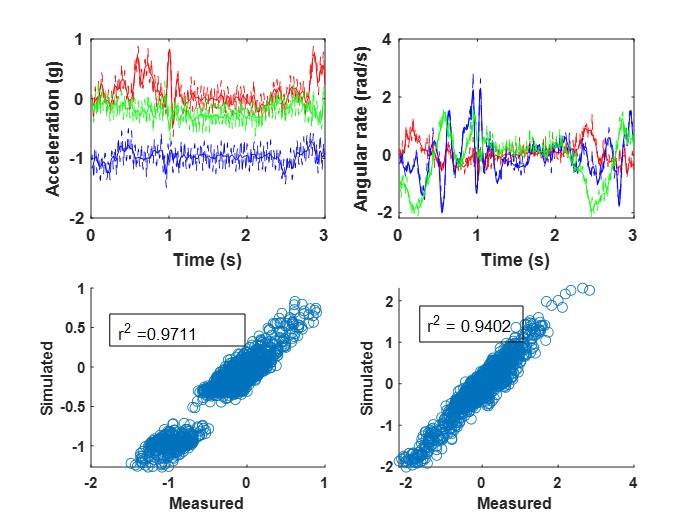}}
\subfigure[Comparison of measured and simulated data with an IMU movement. ]{
\includegraphics[width = 0.4\textwidth]{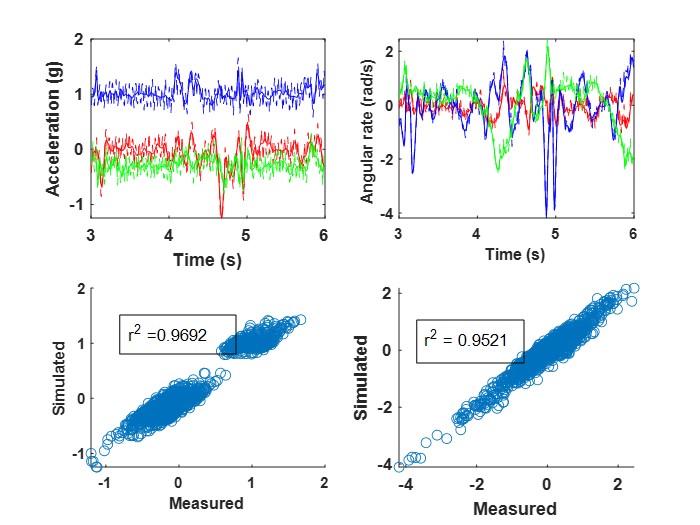}}
\caption{Comparison of measured and simulated data. Solid lines in time series plots are simulated data, faded lines are real IMU measurements.  }
\label{simulated_sig}
\end{figure}

\textbf{Results:} Similar to the evaluation of \cite{young2011imusim}, we also present our results in terms of 3-second curves and correlations. As shown in Fig. \ref{simulated_sig}, the correlations before IMU movements between simulations and measurements are $0.9711$ for acceleration and $0.9402$ for angular rate. The correlations after IMU movements between simulations and measurements are $0.9692$ for acceleration and $0.9521$ for angular rate. 

2. \textbf{Evaluating the mathematical formulation:} We evaluate our mathematical formulation of the metrics by 1) evaluating them under our adopted assumptions and 2) evaluating their variations using our full simulation model. Note that we evaluate the mathematical formulation by our simulation model. For the first purpose, we simulate the 1-DoF condition and calculate the metrics in the sliding windows at the same time duration. The only difference between the two sliding windows is the IMU movement, i.e. one window containing original IMU signals and the other containing moved IMU signals. And the interval of the two windows is 0. Under our adopted assumptions, both windows contain the simulated signals from the same time duration. We calculate $cross(\bm{a}, \bm{j}) = (\sum_t (\bm{ \Delta j_{knee}^{[s_T]}})^T \cdot \bm{ \Delta a_T^{[s_T]} }(t))$, and $ diff(\bm{\omega}, \bm{j}) =
 (\|\bm{\Delta j_{knee}^{[s_T]}} \|_2^2 - \frac{\sum_t \|\bm{ \Delta \omega_T^{[s_T]} }(t) \|_2^2 }{\sum_t \|\bm{ \omega_T^{[s_T]} }(t) \|_2^2}     ) $ in each pair of sliding windows. Note that we use the real value of the difference of the joint axis before and after the IMU movement, i.e. $\bm{\Delta j_{knee}^{[s_T]}}$ rather than $\bm{\Delta \tilde{j}_{knee}^{[s_T]}}$. For the second purpose, we use our full simulation model to synthesize data for real IMU movements. We evaluate the averages, standard deviations (SDs) and histograms of our designed metrics without IMU movement and under IMU movements with various magnitudes. Then, we perform the repeated measures analysis of variance (ANOVA) on the averages and SDs  and plot the histograms. In so doing, we can evaluate the statistical differences of the metrics with and without IMU movement. For plotting the histograms, we select the histograms of  no IMU movement, IMU movement with magnitudes of 1/200 $\cdot \pi$, 1/2$ \cdot \pi$ and $ \pi$ as examples. 

\begin{figure}[ht]
\centering
\subfigure[The averages and standard deviations of the metrics.$*$ denotes the significant difference between with and without IMU movements (ANOVA, p < 0.050).]{
\includegraphics[width = 0.21\textwidth]{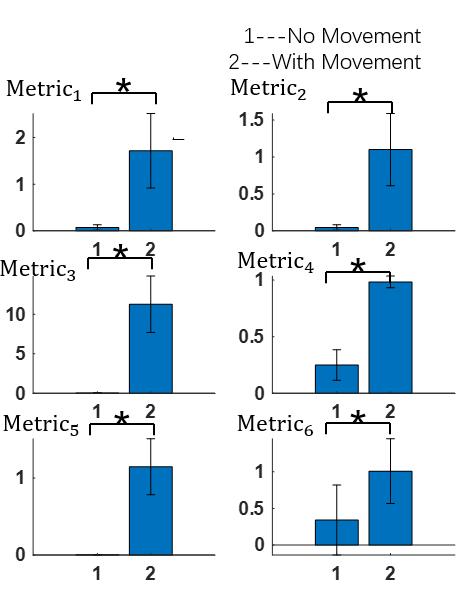}}
\subfigure[The histogram of the metrics under different magnitudes of IMU movements.]{
\includegraphics[width = 0.24\textwidth]{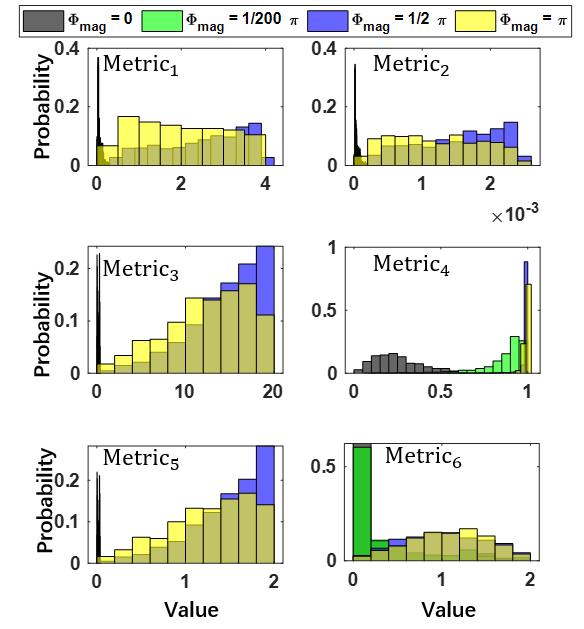}}
\caption{The analysis on the metrics using the data synthesized by our full simulation model.  }
\label{analysis}
\end{figure}

\textbf{Results:} When evaluating for the first purpose, the averages and standard deviations of $cross(\bm{a}, \bm{j})$ and $diff(\bm{\omega}, \bm{j})$ are $2.3e-4 \pm 0.2e-4$ and $1.2e-10 \pm 1e-10$, respectively. As shown in Fig.\ref{analysis} (a), significant differences exist in all the metrics between with and without IMU movement. And the all the metrics present larger values. This would indicate that all the metrics could be used to detect IMU movements. As shown in Fig. \ref{analysis} (b), the histograms of the metrics without IMU movement are difference from those with IMU movements. Moreover, the differences of the histograms increase when the magnitude of IMU movements increases. This would indicate the feasibility of determining the optimal threshold.

3. \textbf{Evaluating the designed metrics:} The thresholds of the metrics determined by Algorithm \ref{threshold} using different $interval$s are presented in Tab. \ref{tab:threshold}. Tab. \ref{tab:threshold}  also presents the minimum magnitude that a metric can detect using the synthetic data and Algorithm \ref{threshold}, denoted by $min_{move}$.  We use the thresholds and the metrics to detect the IMU movements in real-user data. As stated above, we treat No. 1 IMU as the original IMU and treat the rest five IMUs as the moved ones. We randomly generate ten $\tilde{t}$s for each subject, and perform our detection algorithm on the data flow. Then we evaluate the detection performance by the ratio of successful detections and misdetections and the time delay of detections. We denote the ratio of successful detections by $R_{det}$, the ratio of misdetections by $R_{mis}$, the calculation time of each metric the time delay of detections as $\Delta_t$, where $R_{det}$ and $R_{mis}$ are given by:

$R_{det} = $ number of windows when IMU movements are successfully detected / number of windows when an IMU movement occurs

$R_{mis} =$ number of windows when misdetections occur / number of windows of the whole data flow

Moreover, the algorithm is performed on a desktop computer(Intel
i7. 3.4 GHz (Intel, Santa Clara, CA, US), 12 Gb RAM,
windows 10), and we get the calculation time for each metric from MATLAB.

 \begin{table}
  \caption{The Thresholds and Minimal Detected Movement}
  \label{tab:threshold}
  \begin{tabular}{ccccc}
    \toprule
\multirow{2}{*}{Metrics} & \multicolumn{2}{c}{Thresholds} & \multicolumn{2}{c}{$min_{move}$($1/200 \cdot \pi$)}\\
\cmidrule{2-5}
& $ 3000$ & $ 5000$ &  $3000$ & $5000$\\
    \midrule
    Metric$_1$ & 0.481 & 0.495 & 47 & 47\\
	 Metric$_2$ & 2.94e-4 & 3.028e-4 & 48 & 50\\
	 Metric$_3$ & 0.155 & 0.234 & 3  & 6\\
	 Metric$_4$ & 0.531 & 0.822 & 3 & 6\\
	 Metric$_5$ & 0.0188 & 0.0276 & 3 & 6 \\
	 Metric$_6$ & 0.556 & 0.570 & 134 & 143 \\
  \bottomrule
\end{tabular}
\end{table}

\begin{figure}[h]
  \centering
  \includegraphics[width=\linewidth]{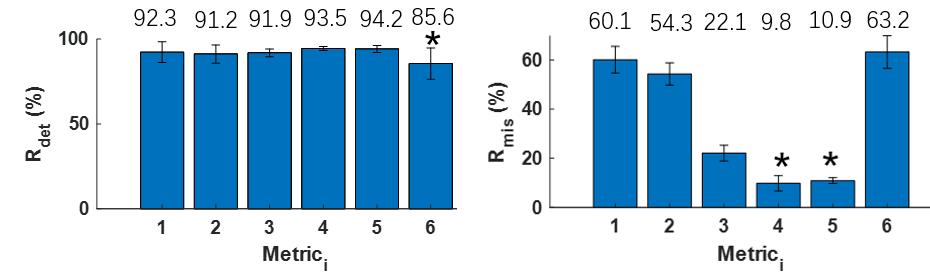}
  \caption{The analysis of successful detection rates and misdetection rates. $*$ denotes the significant difference of the metric compared with other metrics (ANOVA, p < 0.050). In the bottom figure, Metric$_4$ and Metric$_5$ present  significant difference with other metrics, but without difference between them.}
  \label{det_mis_rate}
\end{figure}

\textbf{Results:} As shown in Fig. \ref{det_mis_rate}, all the metrics except for Metric$_6$ present over $90\%$ successful detection rates, and present significant difference with Metric$_6$. Moreover, the misdetection rates of Metric$_4$ and Metric$_5$ are significantly smaller than those of other metrics. It can be seen in Tab. \ref{tab:time} that all the metrics except for Metric$_6$ present similar calculation time. The calculation time of  Metric$_6$  is larger.

 \begin{table}
  \caption{The Calculation Time}
  \label{tab:time}
  \begin{tabular}{cc}
    \toprule
    Metrics & Calculation time (ms)\\
    \midrule
    Metric$_1$ & $2.42 \pm 0.23$ \\
	 Metric$_2$ & $2.56 \pm 0.21$ \\
	 Metric$_3$ & $2.94 \pm 0.12$ \\
	 Metric$_4$ & $2.15 \pm 0.19$ \\
	 Metric$_5$ & $2.66 \pm 0.28$ \\
	 Metric$_6$ & $4.87 \pm 0.11$ \\
  \bottomrule
\end{tabular}
\end{table}

4. \textbf{Evaluating the correction method:}  After detecting the IMU movement, we restart a buffer to collect data and use the data to estimate the coordinates of joint axes, thus re-align the IMU to the body segment. We evaluate the angle estimation errors $before$ the IMU movement and $after$ developing the realignment. 

\begin{figure}[h]
  \centering
  \includegraphics[width=6cm]{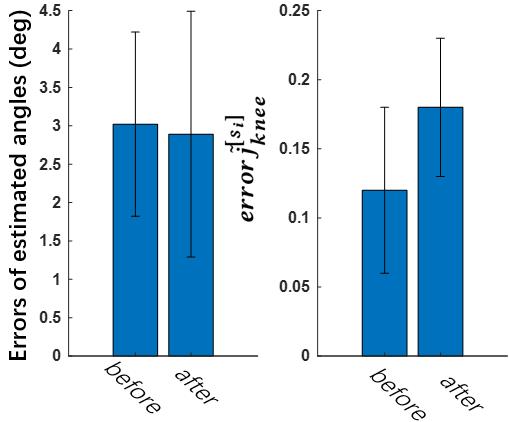}
  \caption{The errors of the estimated angles and joint axis' coordinates $before$ the IMU movement and $after$ developing the realignment.}
  \label{correct}
\end{figure}

\textbf{Results:} The errors of the estimates $before$ the IMU movement and $after$ developing the realignment averaged over subjects are shown in Fig. \ref{correct}. The ANOVA indicates that there is no significant difference between $before$ and $after$, regardless of the estimates. A demo is presented on the website\footnote{https://www.youtube.com/watch?v=PFbHtYrtYy8}.

\section{Discussion}

In this study, we sought to propose an initial step toward online detecting and correcting IMU movements during JAE, with lightweight computations. We first propose a simulation model for IMU movements and integrate it in IMUSim for synthesizing IMU measurements from optical motion capture data. Then, we mathematically formulate how IMU movements affect the IMU measurements and propose metrics accordingly. Based on the synthesized data, we determine optimal thresholds for our proposed metrics. In the evaluation, we demonstrate our simulation model for IMU movements using a similar evaluating paradigm of IMUSim. In addition, we evaluate our proposed metrics and thresholds by real-user data and evaluate the correction method, in terms of accuracy and calculation time.  The results show that Metric$_5$ and Metric$_4$ have better detecting accuracy, lesser misdetection rate can relatively lower calculation time. Moreover, the correction method presents a considerable accuracy of restoring JAE online. These findings indicate that our proposed method is a promising solution to online detecting and correcting IMU movements during JAE. 

The simulation model for IMU movements present a similar accuracy when evaluated in a similar manner as IMUSim. As shown in Fig. \ref{simulated_sig}, the correlations are similar to those presented in \cite{young2011imusim}, when the simulations are evaluated against the IMU movements induced in real-user experiment. This indicate that our simulation model could synthesize IMU measurements under IMU movements with enough validity.

Our designed metrics present different performance on real-user data when using them and the optimal thresholds to detect IMU movements. Metric$_1$, Metric$_2$,Metric$_3$,Metric$_4$ and Metric$_5$ present similar successful detection rates and calculation time. But the misdetection rates of Metric$_4$ and Metric$_5$ are significantly lower than those of Metric$_1$, Metric$_2$ and Metric$_3$. This seems to meet the phenomena of the minimal detected movement presented in Tab.\ref{tab:threshold}, which are determined by synthetic data. This could also indicate Metric$_4$ and Metric$_5$ might be a better selections to detect IMU movements. 

Surprisingly, Metric$_6$, i.e. $\bm{ \Delta \tilde{j}_{knee}^{[s_T]}}$ does not present satisfying results, which might violate our intuitions. This can be attributed to the estimation error of the joint axes’ coordinates. The estimation errors of $\bm{ \tilde{j}_{knee}^{[s_T]}}$ and $\bm{ \tilde{j}_{knee}^{[s_S]}}$ are similar to the errors reported in [2012 Seel]. The errors make the differences of joint axes’ coordinates between sliding windows present a considerable variation, thus might contribute to the lower detection rate. Moreover, the calculation time of calculating  $\bm{ \Delta \tilde{j}_{knee}^{[s_T]}}$ is larger than that of calculating other metrics, since optimizations have to be performed. This makes  $\bm{ \Delta \tilde{j}_{knee}^{[s_T]}}$ a less attractive metric.

The correction method that restarts a buffer and estimates joint axes present considerable accuracy. The accuracy shown in Fig. shows that re-estimating joint axes would not significantly harm the accuracy of JAE. This makes the correction method a promising solution.

\section{Conclusion and Future Work}

In this work, we reveal the issue of IMU movements during JAE  by real-user experiments, mathematically formulate how the IMU movement affects IMU measurements and technically demonstrate our proposed metrics and correction method. The IMU movement detection and correction method we propose is an initial trial and demonstrated to be effective and computationally lightweight. Future work includes the extension to the upper limb joints and trying to accommodate novelty detection algorithms.

\begin{acks}
The experiment protocol was approved by the local ethical
committee and all participants had been informed of the
content and their right to withdraw from the study at any
time, without giving explanation.
\end{acks}

\bibliographystyle{ACM-Reference-Format}
\bibliography{sample-base}

\end{document}